\documentclass[%
reprint,
nofootinbib,
prb,
onecolumn, notitlepage, 11pt
]{revtex4-2}

\usepackage{bm}
\usepackage[utf8]{inputenc}
\usepackage [english] {babel} 
\usepackage{amsmath,amssymb,latexsym}
\usepackage{slashed}
\usepackage{graphicx}
\usepackage{epstopdf}
\usepackage{mathtools}
\usepackage[bottom]{footmisc}
\usepackage{diagbox}

\usepackage{xcolor}
\usepackage{hyperref}

\hypersetup{
  colorlinks=false
  urlbordercolor=gray,
  citebordercolor=gray,
  linkbordercolor=gray,
  pdfborderstyle={/S/U/W 1}
}

\usepackage{amsthm}

\numberwithin{equation}{section}

\newtheorem{assumption}{Assumption}

\def \sec{\begin{section}}
\def \esec{\end{section}}

\def \la {\lambda}

\def \om {\omega}

\def \th {\theta}

\def \Gc {\mathcal{G}}

\def \Lc {\mathcal{L}}

\def \Oc {\mathcal{O}}

\def \Ec {\mathcal{E}}

\def \Ec {\mathcal{E}}

\def \pr {\partial}
\def \ra {\rightarrow}

\def \beq { \begin{equation}}
\def \eeq {\end{equation}}

\def \at {\biggl{\vert}}

\DeclareMathOperator*{\Tr}{Tr}

\renewcommand\Re{\operatorname{Re}}
\renewcommand\Im{\operatorname{Im}}
\newcommand\const{\operatorname{const}}

\renewcommand{\tilde}{\widetilde}
\renewcommand{\bar}{\overline}

\def \l {\left(}
\def \r {\right)}

\def \bra {\langle}
\def \ket {\rangle}

\def \luv {\Lambda_{\rm uv}}

\makeatletter\AtBeginDocument{\let\@elt\relax}\makeatother

\begin{document}

\title{On minimal residual entropy in $0+1$d non-Fermi liquids}

\author{Alexey~Milekhin}
\email{milekhin@ucsb.edu}
\affiliation{University of California Santa Barbara, Physics Department}
\date{\today}
\begin{abstract}
In the large $N$ limit a physical system might acquire a residual entropy at zero temperature even without ground state degeneracy.
 At the same time poles in the 2-point function might coalesce and form a branch cut. Both phenomena
 are related to a high density of states in the large $N$ limit.
 In this short note we address the question:
    does a branch cut in the 2-point function always lead to non-zero residual
    entropy? We argue that for generic fermionic systems in $0+1$ dimensions in the mean-field approximation the answer is positive: branch cut $1/\tau^{2 \Delta}$ in the 2-point function does lead to a lower bound
    $N \log{2}(1/2-\Delta)$ for the entropy. We also comment on higher-dimensional generalizations and relations to the holographic correspondence. 
    \end{abstract}
\maketitle


\section{Motivation}

The large $N$ limit, when the number of interacting fields is taken to infinity, often leads
to nice analytical results. However, the limit itself should be taken with great care. In this paper
we concentrate on two particular features of this limit. In $0+1$ dimensions(quantum mechanics) and
at finite $N$, there is always a finite number of energy levels in the system. Hence any two-point
function at zero temperature has a spectral decomposition as a sum over poles:
\beq
\bra \Oc(\omega) \Oc(0) \ket \sim \sum_n \frac{\mathcal{M}_n}{\omega - \omega_n}.
\eeq
However, in the large $N$ limit the poles can coalesce and form a branch cut $\omega^{2 \Delta-1}$(with some fixed $\Delta$) instead.
Examples of such behavior include Sachdev--Ye--Kitaev(SYK) model \cite{SachdevYe, Sachdev:2015efa, KitaevTalks, PhysRevB.63.134406}, its tensor counterparts \cite{Gurau:2009tw, Gurau:2011aq, Witten:2016iux, Klebanov:2016xxf, Klebanov:2017nlk}, similar Condo models \cite{Sengupta} and
Banks--Fischler--Shenker--Susskind(BFSS) matrix model \cite{Banks:1996vh, Itzhaki:1998dd}. Such branch cut resembles a non-Fermi liquid behavior of higher-dimensional theories.
Another feature of the large $N$ limit is the possibility of non-zero residual entropy at zero
temperature \textit{without} ground state degeneracy. This can happen because the density of states
is exponential in $N$. Examples again include SYK model and its tensor cousins. Specifically, in SYK/tensor models:
\beq
\label{eq:g_syk}
G_{SYK}(\tau) =  \bra \psi_i(\tau) \psi_i(0) \ket \propto \frac{1}{\tau^{2 \Delta}}, \ 1/J \ll \tau,
\eeq
and 
\beq
\label{eq:syk_entropy}
S_{0,SYK}/N = \int_0^{\frac{1}{2} - \Delta} dx\  \frac{\pi x}{\tan{\pi x}}.
\eeq
Both of these phenomena happen because the density of states is very high.  Of course,
we should talk about two different densities: for two-point function
we are interested in the density of "single-particles" excitations, whereas for the entropy we should
talk about the whole spectrum. One needs extra physical input to relate the two.
For example, in SYK it is mean-field approximation. For other theories it might be the requirement
of having a holographic gravity dual.

\textit{The purpose of this short note is to use a mean-field type approximation to argue that 
a branch cut in fermionic two-point function does imply a lower bound on residual entropy.} Specifically,
we consider particle-hole symmetric case and the following branch-cut behavior at zero temperature
\beq
\label{eq:gcut}
G_{\beta=\infty}(\tau) =  \bra \psi_i(\tau) \psi_i(0) \ket \propto \frac{1}{\tau^{2 \Delta}}, \ 1/J \ll \tau,
\eeq
where $J$ is some temperature-independent energy scale. Note that the branch cut extends all the way to $\tau=\infty$ in Euclidean time. This will be important for
our analysis and we discuss this momentarily.

We show that such behavior necessarily implies a lower bound on zero-temperature (generalized) entropy
\footnote{This generalized entropy has nothing to do with black hole generalized entropy.}
$\widetilde{S}_0$:
\beq
\label{eq:bound}
\boxed{\widetilde{S}_0/N \ge \log(2) \l  \frac{1}{2} -  \Delta \r}.
\eeq
 and $\widetilde{S}_0$ is defined as:
\beq
\label{eq:syk_log}
\widetilde{S_0}/N = \lim_{\beta \ra \infty} \sum_{n} \frac{1}{2} \log G_\beta(i \om_n),
\eeq
where the sum is taken over Matsubara frequencies $\om_n$. The bound is saturated for generalized free fields (defined by eq. (\ref{eq:gff})).

Term "generalized entropy" is discussed in detail in Section \ref{sec:log}. We will show that for large $N$ (i.e. mean-field approximation) this quantity is simply
the residual entropy plus certain thermal susceptibilities, eq. (\ref{gen_entropy}). Also in Section \ref{sec:holo}
we discuss why this definition is natural for holographic theories.

\begin{figure}[!ht]
\centering
\includegraphics{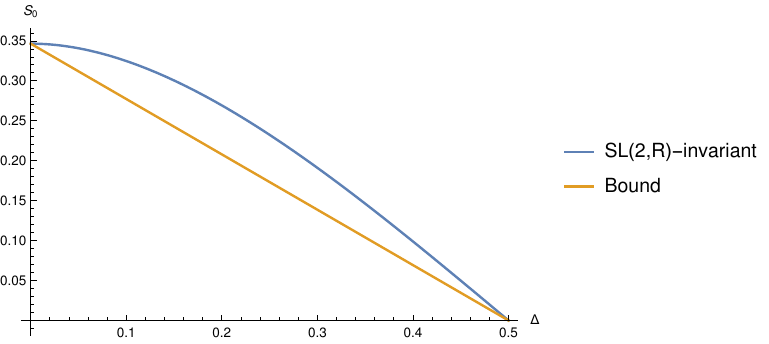}
\caption{Comparison between the conformal $SL(2,R)$-invariant answer (\ref{eq:syk_entropy}), (blue) and the bound (\ref{eq:bound}), (orange).}
\label{fig:comp}
\end{figure}

Notice that the branch cut in the Green function was assumed to be at zero temperature, whereas
in the equation above we have to deal with a finite temperature Green function. So we will need
a few extra assumptions about $T \ra 0$ behavior of the Green function. These assumptions are of dynamical nature and are not related to extra symmetries. Assuming extra symmetries actually make this problem "trivial". Let us briefly explain what we mean.
In $1+1$D CFT the thermal two-point functions in the infinite volume are actually fixed by a conformal symmetry. The form (\ref{eq:gcut}) suggest some type of scale symmetry. However, in $0+1$D the 
situation is more subtle as all mappings are conformal. In the SYK/tensor model case the actual symmetry group is $SL(2,R)$ which does determine finite-temperature Green function uniquely from zero temperature
answer: $G \sim \l \sin(\pi \tau/\beta)\r^{-2 \Delta}$. In turn, it completely fixes $\widetilde{S_0}$ and the answer is given by eq. (\ref{eq:syk_entropy}), the corresponding
computation is identical to the SYK one.
The comparison between $SL(2,R)$ answer (\ref{eq:syk_entropy}) and our bound (\ref{eq:bound}) is presented on Figure \ref{fig:comp}.

Presence of $SL(2,R)$
is something special. For example, in BFSS model one expects to find power-law correlators (\ref{eq:gcut}), but
there is no $SL(2,R)$ symmetry. Also it was recently discussed in \cite{Blake:2017ris, Blake:2018leo, Blake:2021wqj} that the boost symmetry(which is a part of $SL(2,R)$) enforces the theory to be maximally chaotic.
Here we do not want to assume the presence of $SL(2,R)$
symmetry and instead derive a lower bound on $\widetilde{S_0}$.

It is important to emphasize the position of the
branch cut in the two-point function. In eq. (\ref{eq:gcut}) the range of (Euclidean)
time $|\tau|$ is $[1/J,+\infty)$.  We assume that it is valid for any time separation bigger than 
the scale $1/J$. In the frequency domain it results in $\omega^{2 \Delta-1}$ behavior
for any $\omega$ less than $J$. Famously, matrix model resolvents and Green functions
do have a branch cut in the frequency domain, but they do not have residual entropy.
The reason for this, is that the branch cut does not extend all the way to $\omega=0$.
For example in the random-hopping model, 
also known as $q=2$ SYK, 
exact answer for the retarded Green function at finite temperature is
\beq
G_{R,RH}(\omega) \propto \frac{1}{\omega + \sqrt{\omega^2 - 4 J^2}}.
\eeq
For $|\omega| \ll J$, $G_{R,RH}(\omega)$ is simply $\propto 1/J$ and
does not have a branch cut. This model does not have exponentially big in $N$ density of states.

The logic of this paper is following:
\begin{itemize}
    \item 
    In Section \ref{sec:bound} we carefully examine the zero temperature limit of $\Tr \log G$ and derive the
    bound on $\widetilde{S_0}$.

    \item Then in Section \ref{sec:log} we give a general large $N$ argument of why $\Tr \log G$ captures 
    the residual entropy. Also we give a precise relation between the generalized entropy $\widetilde{S}_0$ and standard residual entropy $S_0$.
\end{itemize}

What are the possible caveats in the proposed bound?
Up until now, we implicitly assumed that we are computing the fermionic
contribution to the residual entropy. Surprisingly, in Section \ref{sec:caveats}
we find that a similar bosonic $\Tr \log G$ contributions can 
give a \textit{negative} contribution to the residual entropy.
We will give an explicit example with supersymmetric SYK model \cite{Fu:2016vas}.

A related issue concerns the type of large $N$ limit we study.
As a rule, certain couplings have to be rescaled in the large $N$ limit.
This will be important for us in Section \ref{sec:log} when we discuss
the origin of generalized entropy. However, it is also possible to
have a $N$-dependent \textit{number of couplings}. In this case the analysis
of Section \ref{sec:log} has to be slightly modified and we might need
to include extra bosonic contributions which might lower the entropy,
even if the original theory is purely fermionic. This is illustrated 
in Section \ref{sec:caveats} by supersymmetric SYK model.
Section \ref{sec:disc} is dedicated to a brief discussion of the results and open question.
In Appendix \ref{app:resyk} we compute the residual entropy in
the complex SYK model \cite{KitaevRecent}. We do this to illustrate that one of our technical
assumptions (Assumption \ref{as:T2} , how $G_R(\omega)$ approaches the zero-temperature limit) is important \footnote{Also, to the best of the author's knowledge this is the first \textit{direct} evaluation of residual entropy in this model. The usual route is to relate $S_0$ to a certain charge
susceptibility and then evaluate it. We will demonstrate an UV subtlety in $\Tr \log G$ evaluation.}. 

\section{The bound
from branch-cut}
\label{sec:bound}
This Section is dedicated to studying $\Tr \log G$. We assume that it has a power-law decay at certain frequencies and then derive a useful bound for its low-temperature behavior. In the next Section we explain why it is actually related to the entropy.

Following \cite{Sengupta}, we can rewrite eq. (\ref{eq:syk_log}) in the real frequency domain. We introduce 
the Fermi function $n_F(\om) = 1/(e^{\om \beta}+1)$ and convert the residues into two integrals, right above and right below
the real axis:
\beq
\label{eq:step1}
\frac{1}{2} \sum_n \log G(i \om_n) = \frac{\beta}{4 \pi i} \int_{-\infty}^{+\infty} 
d \om \ n_F(\om) \l \log  \frac{G_R(\om)}{G_A(\om)}  \r.
\eeq

\begin{figure}
    \centering
    \includegraphics{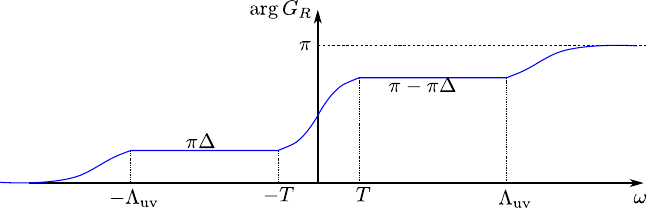}
    \caption{Sketch of $\arg G_R$ as a 
    function of real frequency $\omega$. Regions $|\omega| \gtrsim \Lambda_{\rm uv}$ and $|\omega| \lesssim T$ are not under control.
    Assumption about the power-law form (\ref{eq:conf}) implies that $\arg G_R$ develops two long plateaus.
    In the main text we argue that the region $|\omega| \gtrsim \Lambda_{\rm uv}$ contributes to the ground state energy only.
    }
    \label{fig:argGR}
\end{figure}

Now the phase of the Green function is important. Let us define it as 
\beq
G^> = i \bra \psi_i(t) \psi_i(0) \ket.
\eeq
For Majorana fermions or in the presence of a particle-hole symmetry:
\beq
G^<(t) = -G^>(-t) \ \text{and} \left[G^>(t) \right]^* = -G^>(-t).
\eeq
Hence the retarded and advanced Green functions
\beq
G_R(t) = \theta(t) \l G^>(t) + G^>(-t)  \r,
\eeq
\beq
G_A(t) = - \theta(-t) \l G^>(t) + G^>(-t) \r,
\eeq
are purely imaginary. From that we infer that
\begin{align}
\label{eq:oddeven}
\Re G_R(\om) - \text{odd function}, \nonumber \\
\Im G_R(\om) - \text{even function}. 
\end{align}
Note that it does not imply that $\arg G_R$ is an odd function. Rather, we have a relation:
\beq
\arg G_R(-\omega) = \pi - \arg G_R(\omega).
\eeq
A qualitative sketch of $\arg G_R$ is presented in Figure \ref{fig:argGR}.
Also there is a general relation
\beq
\left[ G_R(\om) \right]^* = G_A(\om).
\eeq
Therefore we can rewrite eq. (\ref{eq:step1}) as 
\beq
\label{eq:intint}
\frac{\beta}{2 \pi } \int_{-\infty}^{+\infty} 
d \om \ n_F(\om) \l \arg G_R(\om) \r.
\eeq

Our main assumption about the Green's function behavior is the following:

\begin{assumption}
\label{as:r1}
There is \textit{temperature-independent} scale $\luv$ below which $G_R$ has a branch-cut
singularity:
\beq
\label{eq:conf}
G_R \l \om \r = \const \frac{i e^{i \pi \l \frac{1}{2}-  \Delta \r}}{\om^{1-2 \Delta}},
\ 1/\beta \ll |\omega| \ll \luv.
\eeq
This is essentially saying that there is a branch-cut at $T=0$ below energy
\footnote{$\luv$ does not have to be a coupling constant, it just sets the UV scale. For example in SYK, we can have several terms in the Hamiltonian:
$
H_{\rm SYK} \propto J_4 \psi^4 + J_6 \psi^6
$
In this case conformal solution $1/\sqrt{\tau}$ is valid upto energies
$\luv = \text{min} \{ J_4 , J_4^3/J_6^2 \}$.}
$\luv$.
The choice of the phase is consistent with eq. (\ref{eq:oddeven}). 
\end{assumption}

Let us return to the computation of the integral
(\ref{eq:intint}).
At very large frequencies $\om \gg \luv$ we are expecting 
free fermion behavior $G_R \sim -1/\om$. So the integral is convergent.
Contribution from $[\luv,+\infty]$ vanishes exponentially for $\beta \ra +\infty$, since 
\beq
\beta \int_{\luv}^{+\infty} d\om \ n_F(\om) = \log \l 1+e^{-\beta \luv} \r.
\eeq
We need to understand if the 
interval  $[-\infty,-\luv]$ gives any $\beta^0$ contribution.
This question is related to how fast does $G_R(\omega)$ approach
its form at zero temperature. If the approach is faster than 
$1/\beta$ when obviously there are no $\beta^0$ contributions. 
Hence our next assumption is
\begin{assumption}
\label{as:T2}
The retarded Green function $G_R(\omega,T)$ at finite temperature approaches its form at zero temperature $G_R(\omega,T=0)$ faster than $T$.
In other words,
\beq
\lim_{T \ra 0} \frac{|G_R(\omega,T)-G_R(\omega,0)|}{T} = 0.
\eeq
\end{assumption}
Roughly this Assumption can be restated as follows: at low energies we have a conformal Green function (\ref{eq:conf}). Behavior
at higher energies is captured by some conformal perturbation theory
with marginal or irrelevant operators. Then we need to require that $\Delta=1$ operator is not present in this perturbation theory.
To demonstrate that this assumption is important, in Appendix  {\ref{app:resyk}} we 
will compute the residual entropy in complex SYK model where this
statement \textit{does not} hold. We will explicitly see that 
the interval $[-\infty,-\luv]$ does give an extra 
contribution\footnote{One underappreciated fact about Majorana
SYK is that $|G_R(\omega,T)-G_R(\omega,0)|$ does go to zero as $T^2$.
This can be seen explicitly at large $q$. At finite $q$ this
statement can be guessed for $|\om| \ll J$ from the expansion 
$\sin^{-2 \Delta}(\pi t T) = t^{-2 \Delta} + \frac{1}{3} \pi^2 t^{2-2\Delta} \Delta T^2 + \Oc(T^4)$ and then checked with numerics
for any $\omega$.}.

So we are left with
\beq
\label{eq:step2}
\frac{\beta}{2 \pi } \int_{-\luv}^{\luv} 
d \om \ n_F(\om) \l \arg G_R(\om) \r.
\eeq
The argument of $G_R(\om)$ goes from $\pi \Delta$ to 
$\pi-\pi \Delta$ - Figure \ref{fig:argGR}.
Equation (\ref{eq:step2}) contains a constant shift to the ground state energy. 
We can see this by noticing
that changing the phase of $G_R$ by $\phi_0$ will change the integral by
\beq
\phi_0 \beta \luv + \Oc(e^{-\beta \luv}).
\eeq
In order to eliminate this shift, 
we need to make sure $\arg G_R(\om) \ra 0$ for $\om \ra -\luv$, since 
$n_F(\om)$ does not decrease in this direction. Thanks to the eq. (\ref{eq:conf}) we know
that this shift is $\pi \Delta$. So we simply need to take the 
limit $\beta \ra +\infty$ in
\beq
\widetilde{S}_0 = \lim_{\beta \ra +\infty} \frac{\beta}{2 \pi } \int_{-\luv}^{\luv} 
d \om \ n_F(\om) \l \arg G_R(\om)  -\pi \Delta \r.
\eeq

\textit{The crucial observation for obtaining the bound is the following.} We can define an odd function of the frequency $\tilde{\arg G_R}$:
\beq
\tilde{\arg G_R} = \arg G_R - \frac{\pi}{2}.
\eeq
Now we need to study
\beq
\label{eq:step3}
\widetilde{S}_0 = \lim_{\beta \ra +\infty} \frac{\beta}{2 \pi } \int_{-\luv}^{\luv} 
d \om \ n_F(\om) \l \tilde{\arg G_R(\om)} + (1-2 \Delta) \frac{\pi}{2}\r.
\eeq

Notice if we substitute $n_F(\om)$ by
$n_F(|\om|)$, the part with $\tilde{\arg G_R(\om)}$ will cancel out(since $\tilde{\arg G_R}$ is odd).
However, this will decrease the value of the integral only if $\tilde{\arg G_R} + \pi (1-2\Delta)/2$ 
is positive everywhere. This is true if the following assumption holds:
\begin{assumption}
\label{as:arg}
$\arg G_R(\omega,T=0)$ is a monotonic function of $\omega \in [-\luv, \luv]$. Or at least 
$\arg G_R- \pi \Delta$ is positive in the interval $[-\Lambda_{\rm uv}, \Lambda_{\rm uv}]$.
\end{assumption}
This assumption can be justified by considering $G_R(\omega)$ at
zero temperature and finite $N$. Its spectral decomposition is given by
\beq
G_R(\omega) = 
\sum_n \frac{ |\bra 0 | \psi_i | n \ket|^2 }{\omega - E_n + i0}.
\eeq
In this form $\arg G_R$ is just a bunch of spikes at $\omega=E_n$ with
their heights determined by $|\bra 0 | \psi_i | n \ket|^2$.
In reasonable physical systems we expect that this matrix element decays
with energy. Therefore, as $N$ grows, the spikes with form a continuous
and monotonic curve\footnote{For example, in Majorana SYK  $\arg G_R \propto \arctan \l \cot(\pi \Delta) \tanh(\om/2)\r$. }. More generally, it is possible to derive a universal bound on the derivative of $\arg G_R$ with respect to the frequency \cite{Zhang:2020jhn}. It would be interesting to use this bound to drop the assumption above.

Finally, we obtain the bound:
\beq
\label{eq:step4}
\widetilde{S}_0 \ge \lim_{\beta \ra +\infty} \frac{\beta}{2 \pi } \int_{-\luv}^{\luv} 
d \om \ n_F(|\om|) \l 1-2 \Delta \r \frac{\pi}{2} = \log(2)\l \frac{1}{2} - \Delta \r.
\eeq
This is the bound for one fermion. 

Let us comment that the bound is saturated if $G_R(\omega)$ is proportional to $\omega^{2 \Delta-1}$ for \textit{all} $\omega$. This is the case of the so-called generalized free fields (GFF). They are characterized by the feature that thermal 2-point function can be obtained via the method of images:
\beq
\label{eq:gff}
G^{GFF}_{\beta \neq \infty}(\tau) = \sum_n (-1)^n G^{GFF}_{\beta = \infty}(\tau + n \beta).
\eeq
Holographic theories typically have this property.

\section{$\Tr \log G$ and (generalized) entropy}
\label{sec:log}
In this Section we relate $\Tr \log G$ to the actual entropy.
The argument below is based on \cite{KitaevRecent} but it appeals to large $N$ only, 
without referring to holography. 

Suppose we have a system of $N$ Majorana fermions without spacial structure: every fermion interacts with every other one. The partition function can be 
written as a path integral in the imaginary time:
\beq
Z_{N} = \int D \psi_i \exp \l -\int d\tau \ \psi_i \pr_\tau \psi_i - H(\psi_1,\psi_2,\dots). \r
\eeq
Let us add another fermion $\psi_0$ to the system. Since the system does not have a spacial structure it is
not ambiguous to do that. 
Now we have
\begin{align}
Z_{N+1} = \int D \psi_i D \psi_0  \exp \Bigg( -\int d\tau \ \psi_i \pr_\tau \psi_i + \psi_0 \pr_\tau \psi_0 - \nonumber \\
-H(\psi_1,\psi_2,\dots) - \psi_0 \xi(\psi_1,\dots) \Bigg).
\label{eq:extra}
\end{align}
Upon integrating out the original $N$ fermions we get a non-local action for $\psi_0$
\begin{align}
Z_{N+1} = Z_N \int D \psi_0 \exp \Bigg( \int d\tau_1 d\tau_2 \ \psi_0(\tau_1) G^{-1}(\tau_1-\tau_2) \psi_0(\tau_2) +  \nonumber \\
+\int d\tau_1 \dots d\tau_4 \ K_4(\tau_1,\dots,\tau_4) \psi_0(\tau_1) \psi_0(\tau_2) \psi_0(\tau_3) \psi_0(\tau_4) + \dots \Bigg)
\end{align}
So far we have not used large $N$ anywhere. Now is the time to use it. In the large $N$ limit, interaction terms like $K_4$
will give a connected contributions to $n-$point correlation functions. We are expecting that they are suppressed in the large $N$ limit
compared to the disconnected contribution. 

The addition of a single fermion produces
\beq
 \frac{\pr \log Z}{\pr N} = \frac{1}{2}\Tr \log G.
\eeq

However, this is not the end of the story, as taking the large $N$ limit almost always involves
tuning some of the couplings. We assume that the couplings we tune are $J_1,J_2,\dots$. 
Hence the full answer for the derivative with respect to $N$ is
\beq
\label{eq:sum}
\frac{d \log Z}{dN} = \frac{1}{2} \Tr \log G + \sum_\alpha \frac{\pr \log Z}{\pr J_\alpha}
\frac{\pr J_\alpha}{\pr N}.
\eeq
In the large $N$ limit we expect that $\log Z = -\beta N F$, hence the left hand side
is exactly the free energy per fermion. At low temperatures
\beq
F = E_0 - T S_0 + \Oc(T^2).
\eeq
 In the right hand side we see the desired 
$\Tr \log G$ term, but we also have extra terms. 
Assuming that the Hamiltonian has the form
\beq
H = H_0 + \sum_\alpha J_\alpha \Oc_{\alpha},
\eeq
we can rewrite eq. (\ref{eq:sum}) as
\beq
\frac{d \log Z}{dN} = \frac{1}{2} \Tr \log G -\beta \sum_\alpha \bra  \Oc_\alpha \ket_\beta 
\frac{\pr J_\alpha}{\pr N}.
\eeq
Then $T^0$ term in the $\Tr \log G$ will compute
the following expression
\beq
\label{gen_entropy}
\boxed{
\widetilde{S_0}=S_0 + \sum_\alpha \frac{\pr J_\alpha}{\pr N}
\frac{\pr \bra \Oc_\alpha \ket}{\pr T} \at_{T=0}}.
\eeq
We will call this quantity \textit{generalized entropy}. 

There is one interesting special case: suppose that the only coupling which gets rescaled with $N$
is $J$ which is also the only dimensionful coupling in the theory. Then it is obvious that
$S_0$ does not depend on $J$ and $S_0$ can be extracted from the $\Tr \log G$ term.
This is the case of the original SYK model. Interestingly, it is also true for the 
BFSS matrix model. However, BFSS model does not have zero temperature entropy.

\section{Possible caveats: $N$-dependent number of couplings and bosons}
\label{sec:caveats}
In this Section we describe the situations when the proposed bound (\ref{eq:bound}) might be violated. So far we have considered only the contribution from fermions. Interestingly, bosons with a branch cut in their
2-point function can contribute \textit{negatively} to entropy. 
The physical reason for this is that if the fermions can form 
a boson, it can condensate and lower the entropy
\footnote{The author is grateful to Leon~Balents for suggesting this picture. Similarly, it was observed in \cite{Romatschke:2019mjm,2021} that dressing photon degrees of freedom can lower the entropy.}.

Also,  notice that in Section \ref{sec:log} we have taken into account the possibility of $N$-dependent coupling \textit{strength}, but not $N$-dependent
\textit{number} of couplings.
As we will see in a moment, such situation can lead to extra bosons, thus 
lowering the entropy.
Let us start from such example.

\subsection{$N$-dependent number of couplings}
Let us say more precisely what we mean by the $N$-dependent number of couplings.
This correspond to the situation, when increasing $N$ by one(adding one extra fermion) leads to an extra term involving $N$ "old" fermions only:
\beq
\label{eq:extra}
Z_{N+1} = \int D \psi_i D \psi_0  \exp \l -\int d\tau \ \psi_i \pr_\tau \psi_i + \psi_0 \pr_\tau \psi_0 - H(\psi_1,\psi_2,\dots) - \psi_0 \xi(\psi_1,\dots) -
\eta(\psi_1,\dots,\psi_N) \r.
\eeq
One such example is $\mathcal{N}=1$ supersymmetric SYK model \cite{Fu:2016vas}. Its Hamiltonian
reads as
\beq
\Lc_{SUSY\ SYK}=\sum_{i=1}^N \psi_i \pr_\tau \psi_i +  \sum_{ijlmn=1}^N C_{ij}^l C^l_{m n} \psi^i \psi^j \psi^m \psi^n,
\eeq
where $C^l_{ij}$ has three indices and each of them goes from $1$ to $N$.
Moreover its components are independent identically distributed Gaussian variables. 
Increasing $N$ to $N+1$ leads to an extra term
\beq
\eta(\psi_1,\dots,\psi_N) =\sum_{1 \le ijmn \le N} C^{0}_{ij} C^{0}_{mn}
\psi^i \psi^j \psi^m \psi^n.
\eeq
This type of interaction can be split into halves by introducing
an extra bosonic field $b^l$:
\beq
\Lc_{SUSY\ SYK} =\sum_i \psi_i \pr_\tau \psi_i - \sum_l \frac{1}{2} b_l^2
+\sum_{ijl} C^l_{ij} b_l \psi_i \psi_j.
\eeq
Now the number of couplings is not $N$-dependent in the above sense. 
As we will show now, such emergent bosons are capable of decreasing the entropy.

\subsection{Bosons}
Following the logic of Section \ref{sec:log}, the contribution to the generalized entropy from a boson can be captured by the determinant:
\beq
\Delta \tilde{S}_0 = - \frac{1}{2} \Tr \log G_b = 
-\frac{\beta}{2 \pi} \mathcal{P} \int_{-\infty}^{+\infty} d\om \ n_B(\om) \arg G_b(\om).
\eeq
Interestingly, this can contribute \textit{negatively} to $\tilde{S_0}$.
Specifically, we assume a similar non-Fermi liquid like behavior plus $SL(2,R)$ symmetry:
\beq
G_{b,SL(2,R)}(\tau) \sim  \l \frac{1}{\sin \l \pi \tau/\beta \r} \r^{2 \Delta_b},
\eeq
then
\beq
\arg G_{b,SL(2,R)} = \arctan(\cot(\Delta_b \pi) \tanh(\beta \om/2)) + \const.
\eeq
For SUSY SYK $\Delta_b = 2/3$. In order to remove the contribution to the ground state energy one needs to shift the phase such that the integral is convergent. In order to compute $\mathcal{P}$ integral one can symmetrize with respect to $\om$. This way one obtains
\beq
\Delta \tilde{S}_{0,SL(2,R)} = \frac{1}{4 \pi} \int_{-\infty}^{+\infty} d \om \Bigg[ \arctan \l \cot \l \Delta_b \pi \r \tanh(\omega/2) \r \coth(\om/2) - \arctan \l \cot(\pi \Delta_b) \r \Bigg].
\eeq
It is easy to compute this integral numerically can see that in the range $1/2 < \Delta_b < 1$ it is always negative. For example, for SUSY SYK it is $-0.0161376$.

In the fermionic case it was possible to obtain the bound simply from asymptotic
of $\arg G$. Here it is not the case. For example, consider the following model
behavior:
\beq
\arg G_{b} = \arctan(\cot(\Delta_b \pi) \tanh(a \beta \om/2)) + \const.
\eeq
By varying parameter $a$ it is possible to change the entropy by arbitrary amount. However, the above example corresponds to unphysical Green's function
$\l 1/\sin \l  a^{-1} \pi \tau/\beta \r \r^{2 \Delta_b}$ which does not
satisfy KMS condition. So maybe it is possible to obtain the bound in the bosonic case, but it will require more sophisticated analysis which includes KMS condition.

\section{A comment on holographic theories}
\label{sec:holo}
\begin{figure}
    \centering
    \includegraphics[scale=1.2]{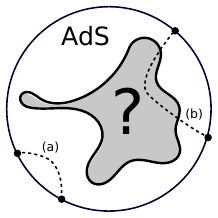}
    \caption{Boundary two point function $(a)$ probes the geometry in the near-AdS region. Whereas $(b)$ probes the geometry deeper inside, the geometry there is denoted by "gray" area and "?".  }
    \label{fig:ads}
\end{figure}

In holographic correspondence \cite{Maldacena:1997re} we expect that at large $N$ some (boundary) theories
  are dual to weakly coupled gravity in higher dimension with Anti-de Sitter (AdS) boundary conditions. From this standpoint the behavior $G_{\rm boundary} \sim 1/\tau^{2 \Delta}$ at short times is typical for holographic states: small time separations probe the geometry in the near-AdS region where one expects the conformal answer. Large time separations probe the geometry further where the answer might be complicated. This is illustrated by Figure \ref{fig:ads}. The bound we derived is not sensitive to the geometry deep inside.

The interaction strength in the bulk
in governed by $1/N$. Therefore from holographic point of view, the thermodynamics at large $N$ should be determined by bulk geometry(like area of the horizons) plus 1-loop determinants
of matter:
\beq
\label{eq:logG_bulk}
-\beta F = S_{\rm cl \ gravity}  + \Tr \log G_{bulk} + \Oc(1/N).
\eeq
Therefore, in holography we directly arrive at the statement that the free energy contains $\Tr \log G_{\rm bulk}$ term.

However, this is not a full story in the holographic case.
Typically in holographic theories the number of bulk fields does not scale with $N$, so the determinant part is subleading compared to the large Bekenstein--Hawking entropy of black hole horizon(which does scale with $N$).

Moreover, eq. (\ref{eq:logG_bulk}) instructs us to consider the Green's function in the dual bulk geometry, instead of the boundary ones.
For free fields in $AdS$ these $\Tr \log$ are not equal. In general, bulk matter can be quantized with two boundary conditions(Dirichlet and Neumann) and the following relation holds  \cite{Hartman:2006dy}:
\beq
\Tr \log G^{\rm Dirichlet}_{\rm bulk} - \Tr \log G^{\rm Neumann}_{\rm bulk} = 
\Tr \log G_{\rm boundary}.
\eeq
So the result we obtained in this paper is not about a single bulk theory, but rather the difference in entropies between the Dirichlet and Neumann quantizations. It would be interesting to extend the results of this paper to bulk Green functions.

It is worth discussing $N$ dependence.
In BFSS
\footnote{Strictly speaking, our results are not applicable for BFSS, as there one should compute $\Tr \log G^{\rm Dirichlet}_{\rm bulk}$ rather than $\Tr \log G_{\rm boundary}$.}
there are actually a lot of operators having a power-law decay 
\beq
\bra \Oc(t) \Oc(t') \ket \propto \frac{1}{|t-t'|^{2\nu+1}}, \ \la^{-1/3} \ll 
|t-t'| \ll \la^{-1/3} N^{10/21},
\eeq
and so having the branch-cuts in 2-point functions\cite{Sekino:1999av,Sekino:2000mg,Kanitscheider:2008kd,Hanada:2009ne}. 
However, they are $SU(N)$ singlets and their number does not scale with $N$.
Correspondingly, BFSS model is not expected to have a residual entropy proportional to $N$. 
There is a version of BFSS model where $SU(N)$ symmetry is ungauged \cite{to_gauge} so one can ask questions about the correlation function of non-singlet operators. However, it has been argued \cite{to_gauge,Numeric} that such operators do not have a power-law decay (non-singlets are gapped). 
Correspondingly is there is no residual entropy even in the ungauged theory \cite{Numeric}.

\section{Discussion}
\label{sec:disc}
In this paper we argued that a branch-cut $\omega^{2 \Delta-1}$ extending all the way to $\omega=0$ in a fermionic two-point function at zero temperature puts a lower bound
on residual entropy(or generalized entropy more generally). We used a few assumptions. The main one being of a mean-field type to justify the extraction of the entropy from $\Tr \log G$. Also we needed a technical Assumption \ref{as:T2} to control how fast the thermal
Green function approaches zero temperature limit. This assumption is violated in
the complex SYK which results in a different residual entropy, which we explain in Appendix \ref{app:resyk}. In conventional quantum field theories it might be challenging to
estimate how $G$ approaches $T=0$ limit, but in holographic theories this is simply
controlled by how fast the spacetime metric approaches $T=0$ limit. In this case the bound is still useful, as computing the whole matter determinant in a curved background is generally very difficult. Another input we needed is Assumption \ref{as:arg} about the behavior of $\arg G_R$. In the main text we briefly
commented that this property is related to how fast certain matrix elements decay with energy. It would be very interesting to make these arguments more sharp.

Finally, let us comment on possible higher-dimensional generalizations. We again can try to look
at the $\Tr \log G$ term. In this case
$\Tr \log$ obviously contains a sum over spacial momenta:
\beq
\label{eq:psum}
\Tr \log G \propto \sum_{n,p} \log G(i \om_n , p).
\eeq
\textit{If} there is a branch-cut $\omega^{2 \Delta-1}$ in the frequency domain for \textit{all} 
spacial momenta $p$, then the results of this paper naturally lead a lower bound on residual
entropy density. In other words, to extensive residual entropy. Unfortunately, the only such theory known to the author is a family of 
SYK chains. Chiral SYK \cite{Lian:2019axs} does not have this property. 
Another interesting 2d analogue of SYK \cite{Turiaci:2017zwd} has non-standard fermion kinetic terms and so the results of this paper are not applicable. Reference \cite{Chowdhury:2021qpy} discusses the possibility \footnote{We thank S.~Sachdev for pointing this out.} of sub-extensive residual entropy. This is realizable in our framework: if for some fraction of momenta $p$ in the sum (\ref{eq:psum}) there is a branch cut, then we can easily obtain a lower bound.
The resulting lower bound will be sub-extensive in volume with the exact volume-dependence determined by the corresponding fraction of momenta with branch cuts. Presumably, even for very small momenta the branch cut will terminate at $\omega \sim p^\alpha, \alpha=\const$. However, if $p$ itself is some inverse power of the volume then the arguments in this paper might be applicable.  We leave the detailed study of this possibility for future work.

\section*{Acknowledgment}
The author is grateful to L.~Balents for raising the question of this paper and providing useful
comments.
Also I indebted to G.~Tarnopolsky for numerous discussions.
It is a pleasure to thank D.~Calugaru, L.~Delacretaz, X.~Dong, A.~Dymarsky, A.~Gorsky, A.~Kamenev, A.~Kitaev, H.~Lin, C.~Liu, D.~Marolf, M.~Mezei, G.~Remmen, S.~Sachdev, U.~Seifert, G.~Turiaci for comments. Also I would like to thank C.~King for moral support.
The work was supported by the Air Force Office of Scientific Research under award number
FA9550-19-1-0360. The work was also supported in part by funds from the University of
California.

\appendix

\section{UV piece in complex SYK}
\label{app:resyk}
In complex SYK model in the presence of the chemical potential
$\mu$ it is more convenient to fix the spectral asymmetry parameter $\Ec$ which is
related to the chemical potential as
\beq
\mu = \mu_0 + 2\pi T \Ec.
\eeq
Hence, the grand canonical thermodynamic potential $\Omega$ is defined by
\beq
d\Omega = (S - 2 \pi \Ec Q) dT - 2 \pi T Q d\Ec.
\eeq
Therefore we will compute $\Gc = S - 2 \pi \Ec Q$ instead of $S$.

Usually $\Gc$ at zero temperature, $\Gc_0 = \Gc(T=0)$, is obtained by first obtaining the following
answer for the derivative \cite{Sengupta}:
\beq
\frac{d \Gc_0}{d \Ec} = - 2 \pi Q.
\eeq
In \cite{KitaevRecent} it was argued that the answer for the entropy can be
obtained by directly computing the $\Tr \log$. However, the actual calculation 
in \cite{KitaevRecent} lifted this $\Tr \log$ to $AdS_2$ space with an electric field 
which made the computation quite involved. In this Section we perform a simple direct (boundary) computation
of the entropy.

Let us start from specifying the details of the model. We will list only the
properties we need for the entropy calculation. We refer to \cite{KitaevRecent} for
a detailed discussion.

The Hamiltonian of the simplest complex SYK model is given by\footnote{We are omitting
bilinear terms which make the Hamiltonian particle-hole symmetric. These terms do not matter
in the large $N$ limit. }
\beq
H_{\rm complex\ SYK} = 
\sum_{j_1< \dots <j_{q/2}, k_1 < \dots <k_{q/2}} J_{j_1 \dots j_{q/2}; k_1 \dots k_{q/2}} 
\bar{\psi}_{j_1} \dots \bar{\psi}_{j_{q/2}} \psi_{k_1} \dots \psi_{k_{q/2}}.
\eeq
At low energies the model has the following Matsubara two-point function in imaginary time
\beq
G(\tau) = - \bra T \psi_j(\tau) \bar{\psi}_j(0)  \ket  = 
c_1 \exp \l 2 \pi \Ec \l \frac{1}{2} - \frac{\tau}{\beta} \r \r 
\l \sin \l \frac{\pi \tau}{\beta} \r \r^{-2 \Delta},
\eeq
and in Euclidean frequencies $\om_n = \frac{2 \pi}{\beta} \l n + \frac{1}{2} \r$:
\beq
\label{G_complex_om}
G(\pm i \om_n) =  \mp i c_2 e^{\mp i \th} 
\Gamma \l \frac{\beta \om_n}{2 \pi} + \Delta \pm i\Ec \r 
\Gamma \l -\frac{\beta \om_n}{2 \pi} + \Delta \mp i\Ec \r \sin \l 
\frac{\beta \om_n}{2} + \pi \Delta \pm i \pi \Ec \r,
\eeq
where 
\beq
\Delta = \frac{1}{q},
\eeq
and $c_{1,2}$ are positive real constants.

Notice that for non-zero $\Ec$, the Green function does not approach its zero-temperature limit faster than $T$:
\begin{align}
\exp \l 2 \pi \Ec \l \frac{1}{2}-\frac{\tau}{\beta} \r \r \sin^{-2 \Delta}(\pi t T) = \nonumber \\
=e^{\Ec \pi }\tau^{-2 \Delta} -2 \pi \Ec e^{\Ec \pi} \tau^{1-2\Delta }T + \frac{1}{3} e^{\Ec \pi} \pi^2 t^{2-2\Delta}(6 \Ec^2+ \Delta) T^2 + \Oc(T^3).
\end{align}

Looking at eq. (\ref{G_complex_om}) we see that $\Ec$ simply shifts the frequency in the real 
domain, such that the retarded Green's function is given by 
\beq
\label{sykc:gr}
G_R(\om) =  - i c_2 e^{- i \th} 
\Gamma \l i\frac{\beta \om}{2 \pi} + \Delta +  i\Ec \r 
\Gamma \l -i\frac{\beta \om}{2 \pi} + \Delta - i\Ec \r \sin \l 
i \frac{\beta \om}{2} + \pi \Delta + i \pi \Ec \r,
\eeq
and its phase is
\beq
\label{G_complex_R}
\arg G_R = -\th + \arctan \l \cot(\pi \Delta) \tanh \l \beta \om/2 + \pi \Ec \r \r.
\eeq
The last piece of information we will need is the relation between $\th$ and $\Ec$:
\beq
\label{th_def}
e^{-2 i \th} = \frac{\cos \l \pi \Delta + i \pi \Ec\r}{\cos \l \pi \Delta - i \pi \Ec  \r}.
\eeq 

Now we have enough information to compute the entropy via $\Tr \log G$.
First of all, 
now we have $G = \bra \psi \bar{\psi} \ket$ and $\bar{G} = \bra \bar{\psi} \psi \ket$
which differ by $\Ec \ra -\Ec$. But we still have the relation $G_R^* = G_A$, hence
in the real frequency domain we have
\beq
\label{q:start}
\frac{\beta}{2\pi} \int_{-\infty}^{\infty} d \om \ n_F(\om) 
\l \arg G_R(\om,\Ec)  + \phi_0 \r + [ \Ec \ra -\Ec ],
\eeq
where the constant shift $\phi_0$ is needed to make 
the integral convergent at $\om \ra -\infty$.

Again, the integral can be separated into two parts: UV part $|\om| \gtrsim \luv$ 
where
Green's functions can be approximated by their non-interacting form and IR part $|\om| \lesssim
\luv$ where we can use the conformal answer.
UV part does not contribute. However, what is $\luv$ in our case? 

In the particle-hole symmetric case(for example, in the original
SYK model with quartic interaction only), $\luv \propto J_4$. 
However, as can be directly seen from eq. (\ref{G_complex_R}), the spectral asymmetry
parameter $\Ec$ acts by shifting the frequency. Therefore physically the UV cut-off
depends on $\Ec$. We can trust the conformal expression (\ref{G_complex_R}) 
in the frequency range 
$[-\tilde{\luv} + 2 \pi \Ec/\beta, \tilde{\luv} - 2 \pi \Ec/\beta ]$, where $\tilde{\luv}$
is some fixed scale which is $\beta$ and $\Ec$ independent. We see that 
the actual UV cut-off becomes $\Ec$ and $\beta$ dependent.

The above discussion is important only for large negative $\om$ where the integral 
in not suppressed by $n_F(\om)$. As follows from the above discussion, $G_R(\om, \Ec)$
and $G_R(\om,-\Ec)$ have different cut-offs and, moreover, the cut-off has $1/\beta$ piece.
Previously we argued that the shift $\phi_0$ is responsible for the ground state
energy. Here it does contribute to the entropy. Specifically, notice that the phase shift $\th$ is $\Ec$-odd according to eq. (\ref{th_def}). So it will not cancel out after adding $-\Ec$ contribution.
We conclude
that the entropy \textit{does receive} the contribution 
\beq
\label{s_uv}
\Gc_{\rm UV} = + 2 \Ec \th,
\eeq
from the UV region because the UV cut-off is IR sensitive.

The rest of the entropy comes from IR region and can be computed as
\beq
\Gc_{\rm IR} = \int_{-\infty}^{+\infty} \frac{d \om}{2 \pi} 
\frac{\arctan \l \cot(\pi \Delta) \tan \l \om/2 + \pi \Ec \r \r + (1-2 \Delta) \pi/2}{e^{\om}+1}
+ [\Ec \ra -\Ec].
\eeq
To check that we have obtained the correct answer we can compute the derivative:
\beq
\frac{\pr \Gc_{\rm UV}}{\pr \Ec} = 2 \th + \Ec \pi \l \tan(\pi \Delta - i \pi \Ec) + 
\tan (\pi \Delta + i \pi \Ec) \r,
\eeq
\beq
\frac{\pr \Gc_{\rm IR}}{\pr \Ec} = 
- \int_{-\infty}^{+\infty} \frac{d\om}{2 \pi} 
\frac{\pi \sin( 2 \pi \Delta)}{\cos(2 \pi \Delta)-\cosh(2 \pi \Ec + \om)} \frac{1}{e^\om+1} 
+ [\Ec \ra -\Ec].
\eeq
The last integral can be computed explicitly. Combining everything together, we get
\beq
\frac{\pr \Gc_0}{\pr \Ec} = \frac{\pr}{\pr \Ec} \l \Gc_{\rm UV} + \Gc_{\rm IR} \r = 
2 \th  - i \pi \l \frac{1}{2} - \Delta \r \l 
\tan(\pi \Delta + i \pi \Ec) -
\tan (\pi \Delta - i \pi \Ec)  \r.
\eeq
This is the correct expression for the derivative \cite{KitaevRecent}.

\bibliographystyle{unsrtnat}
\bibliography{thesis}

\end{document}